# On the theory of superfluidity

Yongle Yu
*State Key Laboratory of Magnetic Resonance and Atomic and Molecular Physics,*
*Wuhan Institute of Physics and Mathematics, Chinese Academy of Sciences, Wuhan 430071, P. R. China*

We investigate the properties of dispersion spectra of one-dimensional periodic Bose systems with repulsive interparticle interactions. These systems with sufficient large interactions can support metastable supercurrent states, which correspond to the local minima of the dispersion spectra at non-zero momenta. The existence of local minima in the spectra and the energy barriers, which separate the minima, can be explained in terms of Bose exchange symmetry. We extend our study to the case of higher dimensional Bose systems. We suggest that superfluidity could be understood as a Bose exchange effect.

## I. INTRODUCTION

Superfluidity, a fascinating quantum phenomenon in which a system demonstrates frictionless motion, is one of the central interests in condensed matter physics. As a many-body quantum effect, however, an exact understanding of superfluidity had remained elusive for more than six decades after its discovery in liquid $^4$He in 1938 [1]. It has be long suspected that superfluidity has something to do with Bose exchange symmetry: the wavefunction of a Bose system is invariant under the exchange of the coordinates of two particles (see, e.g., [2, 3, 4]). It has been pointed out that, especially in the case of one-dimensional Bose systems [4, 5], if the many-body dispersion spectrum of a system contains local minima at nonzero momenta, the many-body states corresponding these minima are metastable and current-carrying, i.e., they are supercurrents, since the energy barriers which separate the local minima prevent the supercurrents from decaying (see Fig. 1 for example). In this paper, we attempt to further argue that the existence of these minima of the dispersion spectrum is due to Bose exchange symmetry. The paper is organized as following, in Sec. II we discuss a kinetic property of a one-dimensional Bose system with translational symmetry, and its connection with the critical velocity of superfluidity. In Sec. III we perform a perturbation analysis of dispersion spectrum of a Bose system in the weak interaction regime. In Sec. IV we compare the dispersion spectrum of a (spinless) Bose system with the dispersion spectrum of a Bose system in which the particles have some internal degrees of freedom (a spinor Bose system). With this comparison we illustrate the role of exchange symmetry in determining the structure of dispersion spectrum of a strongly interacting Bose system. We briefly extend our study to the case of higher dimensional systems in Sec. V, while the detailed discussions on these systems will be presented elsewhere. The conclusion is given in Sec. VI. To avoid confusions, we restrictively define superfluidity as the property that a system supports persistent current in this paper. We don't attempt to link our theoretical model systems to any realistic systems, since our purpose is to seek general qualitative understanding of superfluidity.

## II. GALILEO INVARIANCE AND CRITICAL VELOCITY OF A SUPERFLUID

We first consider $N$ bosonic particles interacting with repulsive interaction, moving along a ring with radius of $R$, i.e., the system has a one-dimensional periodic geometry. The Hamiltonian has the form

$$H = -\sum_{i=1}^{N} \frac{\hbar^2}{2MR^2} \frac{\partial^2}{\partial \theta_i^2} + 2\pi g \sum_{i<j}^{N} f(|\theta_i - \theta_j|), \quad (1)$$

where $g > 0$ is the interaction strength, $\theta_i$ is the angular coordinate, $M$ is the mass of a particle and $f$ describes the form of the interaction ($f \geq 0$). We only consider short-ranged interaction and zero-ranged interaction.

Three quantities, $\hbar$, $M$ and $R$, suffice to generate a set of units for the physical quantities involved in this paper, which will be used to render all quantities dimensionless. The unit of energy, however, is taken to be $\hbar^2/2MR^2$, where an extra factor of one half is involved for convenience.

Due to the axial symmetry of the Hamiltonian, the total angular momentum is a good quantum number. It is straightforward to show that the many-body spectra and eigen wavefunctions, at two angular moments with one differing from the other by $kN$ where $k$ is an integer, can be related in the following way: giving an eigen wavefunction $\psi(\theta_1, \theta_2, ..., \theta_N)$ with angular momentum $L'$ and with energy $E_\psi$, a mapped wavefunction

$$\phi(\theta_1, \theta_2, ...) = e^{ik\theta_1 + ik\theta_2 + ... + ik\theta_N} \psi(\theta_1, \theta_2, ..., \theta_N) \quad (2)$$

will be an eigen wavefunction with angular momentum

$$L = L' + kN \quad (3)$$

and with energy

$$E_\phi = E_\psi + L^2/N - L'^2/N. \quad (4)$$

The factor in Eq. 2, $e^{ik\theta_1+ik\theta_2+...+ik\theta_N}$, describes center-of-mass motion and conserves the inner structure of $\psi$ in this one-dimensional periodic system. The energy difference between $\psi$ and $\phi$ is independent of interaction, and one might attribute this kinetic property to Galileo invariance (see also [5, 6]).

We examine only the dispersion relation, $E = E(L)$, where $E$ is the lowest energy of the system at given $L$. We refer to $E$ as the "yrast" energy, following the convention in nuclear physics. Generally, the yrast energy at $L = 0$ is the global minimum of the dispersion spectrum, i.e., the ground state energy. Other local minima (if they exist) in the yrast energy at nonzero $L$, i.e., $... > E(L-2) > E(L-1) > E(L) < E(L+1) < E(L+2) < ....$, indicate metastable states carrying persistent currents [4, 5]. The peculiar possible existence of the local dispersion minima at none zero $L$ is of quantum nature and beyond classical understanding.

With the mapping (Eqs. 2-4), one will know the full dispersion spectrum if the part at angular regime $0 \leq L' \leq N$ is giving (see Fig. 1). For the energy difference between two neighboring yrast states, with Eq. 2 one easily obtains the following relations,

$$E(L = L' + kN) - E(L - 1 = L' - 1 + kN)$$
$$= E(L') - E(L'-1) + 2k + (2L'-1)/N \quad (5)$$

and

$$E(L = kN + 1) - E(L - 1 = kN)$$
$$= E(N+1) - E(N) + 2k + 1/N. \quad (6)$$

With large enough $k$, for example, $k > max(E(j)-E(j+1))/2$ $(0 \leq j \leq N)$, the right hand sides of Eq. 5 and Eq. 6 are positive for all $L$. Hence, $E(L-1) < E(L) < E(L+1) < E(L+2) < ... < E(\infty)$ with sufficient large $L$, the dispersion spectrum will not support any persistent currents further. This naturally explains that there is an upper limit of (angular) velocity of a superfluid, beyond which the system is not frictionless any more(see also [5]). It will be shown that the yrast spectrum contains a local minimum at $L = N$, giving sufficiently large $g$, and has a local linear dispersion around this minimum. With the analysis above, it is clear that the slope of the linear spectrum at $L = N$, or simply $E(N-1) - E(N)$, related to the slope of the spectrum at $L = 0$, will determine the critical (angular) velocity of the superfluid.

Landau [7] first related superfluidity of a system with the properties of quasi-particle spectrum of the system. One shall distinguish Landau's quasi-particle spectrum and the many-body dispersion spectrum discussed above. A clear difference between two spectra can be seen if one compares their general behavior at rather large momentum $p$, where the quasiparticle spectrum likely approximates $p^2/2M$ but the dispersion spectrum is close to $p^2/2NM$ with $N$ can be macroscopic (the ground state energy is shifted to be zero).

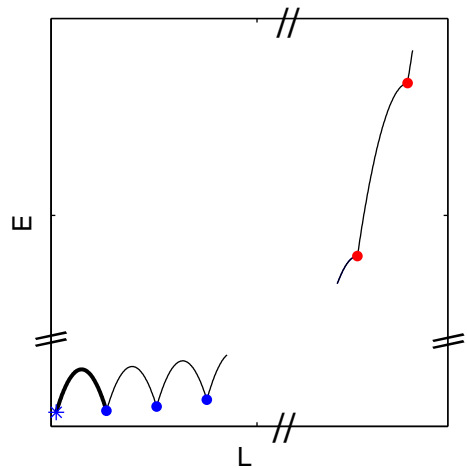

FIG. 1: A schematic plot of many-body dispersion spectrum of a one-dimensional Bose system with repulsive interaction. Once the part of the spectrum at angular momentum regime $0 \leq L \leq N$ (the thick line) is known, the rest of the spectrum can be obtained. The star marks the ground state. filled circles (red or blue) mark the yrast states which can be mapped to the ground state, among which only those with angular momentum less than a certain value are supercurrent states (blue circles). See also [5].

### III. THE STRUCTURE OF THE DISPERSION SPECTRUM AT WEAK INTERACTION REGIME

For the purpose of simplicity and convenience, we shall also use a rotating frame of reference besides the laboratory frame. In a rotating frame with angular velocity $\Omega$, the Hamiltonian is $\hat{H}^\Omega = \hat{H} - \Omega \hat{L}$.

A natural set of single particle orbits can be chosen to be $\varphi_m(\theta) = e^{im\theta}/\sqrt{2\pi}, m = 0, \pm 1, \pm 2, ....$ One easily sees that the yrast states of noninteracting system ($g = 0$) at the angular momentum regime $0 \leq L' \leq N$ are the Fock states, $|0^N 1^0\rangle, |0^{N-1} 1^1\rangle, |0^{N-2} 1^2\rangle, ..., |0^0 1^N\rangle$ (Here $|0^{N_0} 1^{N_1}\rangle$ means that $N_0$ particles occupy the orbit $\varphi_0$, and $N_1$ particles the orbit $\varphi_1$). Only two single-particle states $\varphi_0$ and $\varphi_1$ are involved in these yrast states, while Fock states involving any other single particle orbits are excited states above the dispersion spectrum since they have relatively larger (kinetic) energy. In the rotating frame with $\Omega = 1$, $\varphi_0$ and $\varphi_1$ are the degenerate single particle ground states with zero energy.

In the limit of weak $g$, within the first order perturbation theory, the Fock states $|0^{N-L'} 1^{L'}\rangle (0 \leq L' \leq N)$ remains the yrast states. In the rotating frame with $\Omega = 1$, the kinetic energies of these Fock states are zero. The interaction energy of $|0^{N-L'} 1^{L'}\rangle$, the same as the total energy in the rotating frame, is calculated straightforwardly and can be written in four parts,

$$E^\Omega = (N-L')(N-L'-1)g\epsilon/2 + L'(L'-1)g\epsilon/2$$
$$+ (N-L')L'g\epsilon/2 + (N-L')L'g\epsilon_{ex}/2 \quad (7)$$

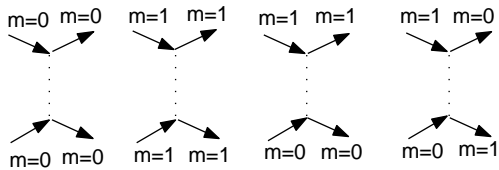

FIG. 2: Interaction diagrams for the Fock state $|0^{N-L'}1^{L'}\rangle$. The dashed line represents the interaction. $m$ is the angular momentum of the single particle orbit. In the last diagram, two particles exchange their orbits.

with $\epsilon = \int_{-\pi}^{\pi} f(|\theta|)d\theta$, and with

$$\begin{aligned}\epsilon_{ex} &= \frac{1}{2\pi}\int_{-\pi}^{\pi}\int_{-\pi}^{\pi} e^{-i\theta_1}e^{-i0\theta_2}f(|\theta_1-\theta_2|)e^{i0\theta_1}e^{i\theta_2}d\theta_1 d\theta_2 \\ &= \int_{-\pi}^{\pi} e^{i\theta}f(|\theta|)d\theta.\end{aligned} \quad (8)$$

One might think that the four terms on the right hand side (RHS) of Eq. 7 correspond to the four interaction diagrams in the Fig. 2, respectively. For example, the first term counts the interaction energy with two (interacting) particles in the orbit with $m=0$. Since there are $N-L'$ particles in this orbit, out of which $(N-L')(N-L'-1)/2$ pairs of particle can be formed. The sum of first three terms on the RHS of Eq. 7 is $N(N-1)g\epsilon/2$ and is independent of $L'$. The last term, causing by the exchange interaction, generates the parabolic energy barrier between the ground state and yrast state at $L=N$ (in the rotating frame). Note that the argument here is equivalent to that of Noziéres in [8] where he relates Bose-Einstein Condensation with exchange interaction.

One might check with a system with $N=2$ to see how the exchange term originates. The wavefunction of the yrast state at $L=0$ is $\Psi_{L=0}(\theta_1,\theta_2) = e^{i0\theta_1}e^{i0\theta_2}/2\pi$, and that at $L=N=2$ is $\Psi_{L=2}(\theta_1,\theta_2) = e^{i\theta_1}e^{i\theta_2}/2\pi$. Both wavefunctions have a product structure and are symmetric under exchange of two Bose particles (since two particles occupy the same orbits).

The wavefunction of the yrast states at $L=1$, is $\Psi_{L=1}(\theta_1,\theta_2) = \sqrt{2}/2(e^{i0\theta_1}e^{i\theta_2} + e^{i\theta_1}e^{i0\theta_2})/2\pi$, which takes a form of a sum of two products, due to the symmetrization. The interaction energy, $V = \int\int \Psi_{L=1}^* 2\pi g f(\theta_1-\theta_2)\Psi_{L=1}d\theta_1 d\theta_2$, second order in the wavefunction, will generate extra cross terms, given that the wavefunction itself is a sum of products instead of a product.

If one ignores the symmetrization and considers an invalid wavefunction at $L=1$, $\Psi_{L=1}^{invalid} = e^{i0\theta_1}e^{i\theta_2}/2\pi$, it will has the same interaction energy as that of $\Psi_{L=0,2}$. Comparing $\Psi_{L=1}$ and the invalid wavefunction, one sees that symmetrization pushes up the (interaction) energy for the yrast state with $L=1$.

## IV. A COMPARISON BETWEEN THE DISPERSION SPECTRUM OF A SPINLESS BOSE SYSTEM AND THAT OF A SPINOR BOSE SYSTEM

In Sec. III, we show that the structure of dispersion spectrum of the Bose system is determined by the exchange interaction in the limit of weak interaction. We shall show that the possible dispersion energy barrier between the yrast state at $L=0$ and that at $L=N$ becomes more apparent at larger $g$, and can be attributed to the exchange interaction. These are illustrated by comparing the dispersion spectra of the Bose system considered above, referred to a spinless Bose system, with the dispersion spectra of a spinor Bose system, which only differs from the spinless system by allowing two degenerate internal states $|\alpha\rangle$ and $|\beta\rangle$ to each particle.

We can formally introduce (pseudo-) spin operators $s^2$, $s_z$ associated with the internal states. For example, $s_z|\alpha\rangle = |\alpha\rangle/2$ and $s_z|\beta\rangle = -|\beta\rangle/2$. We then refer to $|\alpha\rangle$ as spin up and $|\beta\rangle$ as spin down. The interaction does not affect the spin and the spinor Bose system has $SU(2)$ symmetry. In second quantization, the Hamiltonian is,

$$\begin{aligned}H_s &= \sum_{l,\sigma=\alpha,\beta} l^2 c_{l,\sigma}^\dagger c_{l,\sigma} \\ &+ \frac{g}{2}\sum_{k,l,m,n,\sigma,\sigma'} f_{klmn} c_{k,\sigma'}^\dagger c_{l,\sigma}^\dagger c_{m,\sigma} c_{n,\sigma'} \delta_{k+l,m+n}\end{aligned}(9)$$

where $c_{l,\sigma}^\dagger$ ( $c_{l,\sigma}$) is the operator which creates (annihilates) a particle at the single particle state $\varphi_l(\theta)|\sigma\rangle$, and where

$$f_{klmn} = \frac{1}{2\pi}\int\int e^{-ik\theta_1}e^{-il\theta_2}f(|\theta_1-\theta_2|)e^{im\theta_2}e^{in\theta_1}d\theta_1 d\theta_2. \quad (10)$$

Due to the $SU(2)$ symmetry, $S^2 = (s_1+s_2+...+s_N)^2$ and $S_z = s_{z1}+s_{z2}+...+s_{zN}$ are good quantum numbers.

Let us first examine how the spin degrees of freedom interplay with the exchange interaction. A many-body wavefunction of a Bose system is required to be symmetric under the exchange of any two particles. If the wavefunction is symmetric/antisymmetric under the exchange of their spin degrees of freedom, it also has to be symmetric/antisymmetric under the exchange of their spatial degrees of freedom. For example, we consider a system of two Bose particles which occupy different single-particle spatial orbitals, $\varphi_k(\theta)$ and $\varphi_{k+1}(\theta)$. The wavefunction with $S=0$ is

$$|S=0,S_z=0\rangle = \frac{1}{2}[\varphi_k(\theta_1)\varphi_{k+1}(\theta_2) - \varphi_k(\theta_2)\varphi_{k+1}(\theta_1)] \\ \times(|\alpha_1\rangle|\beta_2\rangle - |\alpha_1\rangle|\beta_2\rangle)(11)$$

while the wavefunction with $S=1$ and $S_z=0$ is

$$|S=1,S_z=0\rangle = \frac{1}{2}[\varphi_k(\theta_1)\varphi_{k+1}(\theta_2) + \varphi_k(\theta_2)\varphi_{k+1}(\theta_1)] \\ \times(|\alpha_1\rangle|\beta_2\rangle + |\alpha_1\rangle|\beta_2\rangle)(12)$$

The state with $S = 0$ has an interaction energy of $g\epsilon - g\epsilon_{ex}$ since the exchange energy is equal to $-g\epsilon_{ex}$. For the state with $S = 1$, the exchange energy is $g\epsilon_{ex}$ and the total interaction energy is $g\epsilon + g\epsilon_{ex}$. In this system, the exchange interaction favors the $S = 0$ state, which can be understood easily. With $S = 0$, the spatial part of the wavefunction is antisymmetric and it vanishes with $\theta_1 = \theta_2$, implying that one particle has a small chance to approach and to interact effectively with the other particle.

To demonstrate the spin structure of the $N$-particle yrast states of the spinor Bose system, we will once again discuss the perturbative regime first. At small $g$ and a given $L'$, in first order perturbation the spatial orbital configuration consists of $N - L'$ particles occupying the $\varphi_0$ state and $L'$ particles occupying the $\varphi_1$ state. According to the simple argument given above, it is energetically favorable for the yrast state to accommodate as many spatially antisymmetric pairs as possible. For $L' \leq N/2$, the maximum number of such pairs is $L'$ and the rest $N - 2L'$ particles are in the state $\varphi_0$. Each pair has a zero contribution to the total spin and the unpaired particles are symmetric in spin, thus the total spin is $(N - 2L')/2$. For $N/2 \leq L' \leq N$, $S = (2L' - N)/2$. Our numerical calculation confirmed this spin structure. In addition, we found for the yrast spectrum as a function of $L'$, $E^\Omega = -(N - 2S)g\epsilon_{ex}$, apart from a constant term $N(N-1)g\epsilon/2$ (This constant term will be ignored thereafter). Here $E^\Omega$ takes the form of an exchange energy, where each spatially antisymmetric pair contribute with an exchange energy of $-g\epsilon_{ex}$.

We shall consider the spinor system with any $g$. It is found numerically that the total spin $S$ of a yrast state at a given angular momentum $L'$, as a function of $g$, is stable, i.e., it does not change for any finite $g$. It might be understood in the following way. The exchange interaction plays a dominant role in determining the low energy spectrum. At large $g$, even though the yrast state involves more single-particle spatial orbits and all orbits contribute to the exchange energy, the leading contribution still comes from $\varphi_0$ and $\varphi_1$. To reduce the leading contribution, $S$ will therefore remain the same as in the perturbative regime.

For the spinor Bose system, $E^\Omega(L')$, is linear in $g$ for small $g$, then decreases monotonically with $g$ and finally saturates to $-L'(N-L')/N$ at sufficient large $g$, as we found numerically [9]. Correspondingly, the dispersion relation does not have any local metastable minima in the laboratory frame. Hence, the spinor Bose system does not support persistent currents [10].

We shall compare the spinless Bose system and the spinor Bose system. One can see that a many-body state of the spinor system with $S = S_z = N/2$ is equivalent to a many-body state of the spinless system, since all spins point to the same direction and the spinor system is equivalently spinless. The yrast spectrum for the spinless system corresponds to the lowest energy eigenstate of the spinor Bose system at a given angular momentum, with $S = N/2$ and $S_z = N/2$. The exchange interaction associated with $S = N/2$ pushes up the energy of yrast state of the spinless system, relatively to the energy of the yrast state of the spinor system which has $S = |N - 2L'|/2$. This generates a barrier between the yrast state at $L' = 0$ and that at $L' = N$ in the rotating frame. For $g$ larger than a critical value, the yrast spectrum of the spinless Bose system develops a minimum at $L = N$ in the laboratory frame, allowing for the possibility of metastable persistent currents.

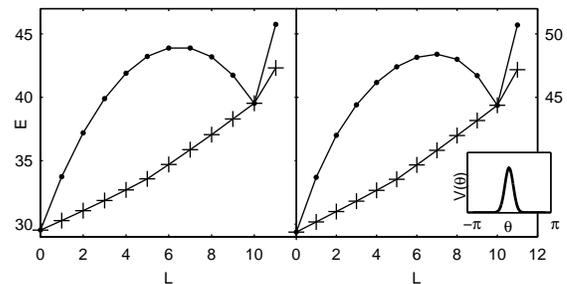

FIG. 3: The yrast spectra in the lab frame. We use dots for the spinless Bose system and crosses for the spinor Bose system. Left panel refers to a contact interaction and right panel to a finite range interaction $f(|\theta|) = e^{-\theta^2/4\epsilon}/\sqrt{4\pi\epsilon}$ with $\epsilon = 0.05$, illustrated in the inset figure. In both graphs, $N = 10$ and $g = 1$.

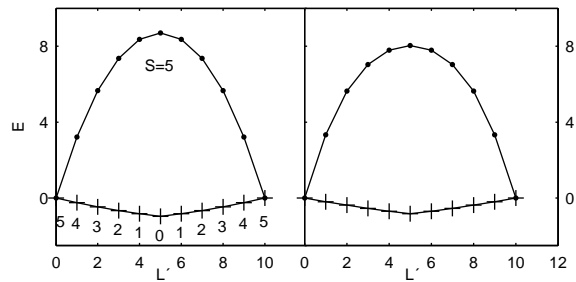

FIG. 4: The yrast spectra in the rotating frame, relative to the yrast energy at $L = 0$. Left panel refers to the delta interaction and right panel to a finite range interaction $f(|\theta|) = e^{-\theta^2/4\epsilon}/\sqrt{4\pi\epsilon}$ with $\epsilon = 0.05$. The total spin $S$ of yrast states is written out in the left panel ($S = N/2$ for spinless system). In both graphs, $N = 10$ and $g = 1$.

As a manifestation of the exchange interaction, for a spinless Bose system $E^\Omega$ increases with the (exchange) interaction strength $g$. This motivates the following relation,

$$E^\Omega(L', g_2) > E^\Omega(L', g_1) \quad \text{if} \quad g_2 > g_1 \qquad (13)$$

At large $N$ and small angular momentum $L' \equiv l > 0$, we assume that $E^\Omega$ has a power series expansion as follows,

$$E^\Omega(l, g) = \alpha_0 + \alpha_1(g)l/N + \alpha_2(g)(l/N)^2 + ... \qquad (14)$$

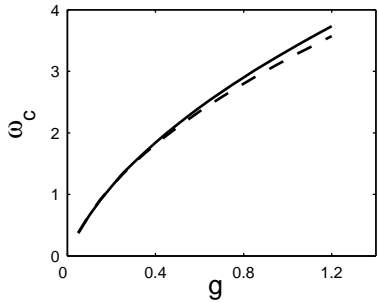

FIG. 5: Critical angular velocity $\omega_c$ as a function of $g$. The dashed line is for the delta interaction, and the solid line for a finite range interaction $f(|\theta|) = e^{-\theta^2/4\epsilon}/\sqrt{4\pi\epsilon}$ with $\epsilon = 0.05$. In both cases $N = 10$.

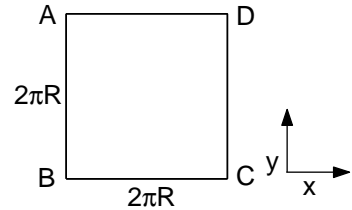

FIG. 6: The geometry of a 2D periodic system. The edge $AB$ is identified with $DC$ and $BC$ identified with $AD$. The linear size of the system is $2\pi R$.

$\alpha_0 = 0$ since $E^\Omega(l = 0, g) \equiv 0$ (the ground state energy is shifted to be zero). In the perturbative regime (small $g$) with $E^\Omega(l,g) = l(N-l)g\epsilon_{ex}$ we find $\alpha_1(g) = gN^2\epsilon_{ex}$ which is not zero. Therefore the linear term is the leading (nonvanishing) term of $E^\Omega(l,g)$. $\alpha_1(g)/N$ might be identified with critical angular velocity $\omega_c$, and thus $\omega_c = gN\epsilon_{ex}$ in the perturbative regime. we ignore all higher terms other than the linear term at the RHS of Eq. 14 at small $l/N$. Combining the two relations 13, 14, we find

$$\alpha_1(g_2)/N > \alpha_1(g_1)/N \quad \text{if} \quad g_2 > g_1 \tag{15}$$

or

$$\omega_c(g_2) > \omega_c(g_1) \quad \text{if} \quad g_2 > g_1. \tag{16}$$

Therefore, the critical angular velocity, as a function of $g$, is linear at small $g$ and increases monotonically with $g$. In order for the system to support a supercurrent at $L = N$, $\omega_c$ needs to be large than unity.

Some numerical examples are plotted in Fig. 3 and Fig. 4. The yrast spectrum of a spinless system and that of the corresponding spinor system are presented in left panel of Fig. 3. For both systems, $N = 10$, $f(|\theta_1 - \theta_2|) = \delta(\theta_1 - \theta_2)$ (,i.e., a contact interaction), and $g = 1$ ($Ng\epsilon_{ex} = 10 \gg 1$, and therefore the interaction is beyond the perturbative regime). It is clear that there is a supercurrent at $L = 10$ for a spinless system while there is no supercurrents for the corresponding spinor Bose system. For both systems the corresponding $E^\Omega$ is presented in the left panel in Fig. 4. The right panels of Fig. 3 and Fig. 4 correspond to the systems with finite range interactions, $f(|\theta|) = e^{-\theta^2/4\epsilon}/\sqrt{4\pi\epsilon}$ with $\epsilon = 0.05$. For both systems, $N = 10$, $g = 1$.

We also calculated the critical velocity $\omega_c$ of spinless systems and plotted $w_c$ vs $g$ in Fig. 5.

## V. HIGHER DIMENSIONAL SYSTEMS

The above understandings of dispersion spectra of one-dimensional Bose systems can be generalized to the case of higher dimensional systems straightforwardly. Here we just briefly discuss 2D periodic systems (see Fig. 6) with finite range interactions while the more detailed discussions and the numerical checks on higher dimensional systems will be presented elsewhere.

The total momentum components $P_x$, $P_y$ of a 2D system are good quantum numbers and we shall consider only the dispersion spectrum $E = E(P_x, P_y)$ where $E$ is the lowest eigen level at a given momentum. Local minima (if they exist) of the dispersion spectrum at non-zero momenta will indicate the supercurrent states.

The perturbative analysis of the dispersion spectra in the limit of weak interaction strength $g$ is straightforward, similar to the case of one-dimensional systems. For example, the yrast states of a spinless system at the momenta $0 \leq P_x \leq N, P_y = 0$ will be approximated well by Fock states $|(0,0)^N(1,0)^0\rangle, |(0,0)^{N-1}(1,0)^1\rangle, |(0,0)^{N-2}(1,0)^2\rangle$, ..., $|(0,0)^0(1,0)^N\rangle$, where $(m,n)$, a pair of integers, denotes the single particle orbit $\psi_{m,n}(x,y) = 1/(2\pi)e^{imx/2\pi R+iny/2\pi R}$. The interaction energies of these states, as a function of $P_x$ will has a parabolic part (the exchange term) $\propto P_x(N - P_x)$ besides a constant part (the Hatree term). If $g$ is above a critical value, the dispersion spectrum will contain a local minimum at $P_x = N, P_y = 0$. By contrast, the dispersion spectrum of a spinor Bose system will contain no local minimum other than the global one at $P_x = 0, P_y = 0$.

Due to the Galileo invariance, the spectrum at large momenta can be written out by mapping them to the spectrum at momenta regime $0 \leq P_x \leq N$ and $0 \leq P_y \leq N$. Following the mapping algebra ( 2D versions of Eqs. 7-9), one can find that the dispersion spectrum of a spinless Bose system will contain no local minimum at given large enough momentum, i.e., there is an upper velocity limit for a superfluid.

One rather new finding about the 2D systems is following: for a spinless system with $N = j^2$ where $j$ is an integer and with given large enough $g$, there will be also supercurrent states at $P_x = k_x j, P_y = k_y j$, where

$k_x, k_y = 1, 2, 3, ..., j-1$. One might note that, with $N = j^2$, the system allows to form a lattice (crystallization) at large repulsive interactions. The new supercurrent states, written as superpositions of many-body Fock states, can not involve any BEC-like states (Fock states with all $N$ particles occupy the same orbit). The origin of the new supercurrent states, can be understood in terms of Bose exchange symmetry (the explanation is presented elsewhere).

## VI. CONCLUSION

We have studied the properties of dispersion spectra of Bose systems with repulsive interactions. We argued that the possible existence of local minima of the spectra and the energy barriers separating these minima are due to the Bose exchange symmetry. Although the numerical checks can be only done to systems with limited $N$, the microscopic physical arguments can be applied to systems with any large $N$.

The quantum phenomenon of superfluidity is rather puzzling. However, one might get a simple picture of superfluidity if one, i) relates it with dispersion spectra of the superfluids [4, 5], ii) explains the non-classical features of the dispersion spectra in terms of the exchange properties of Bose particles. One might draw an analog between this picture of superfluidity and Heisenberg's exchange theory of ferromagnetism [11]. In fact, part of arguments presented in Sec. IV bears a large similarity, both formally and in content, to Heisenberg's arguments on ferromagnetism. In conclusion, we suggest that superfluidity might be understood as a Bose exchange effect.